\documentclass[conference]{IEEEtran}
\IEEEoverridecommandlockouts
\usepackage{cite}
\usepackage{amsmath,amssymb,amsfonts}
\usepackage{algorithmic}
\usepackage{graphicx}
\usepackage{textcomp}
\usepackage{xcolor}
\usepackage{ulem}
\usepackage{enumitem}
\usepackage{bbm}
\usepackage{multirow}
\usepackage{bbding}
\usepackage{url}

\def\BibTeX{{\rm B\kern-.05em{\sc i\kern-.025em b}\kern-.08em
    T\kern-.1667em\lower.7ex\hbox{E}\kern-.125emX}}

\newcommand{\Uset}{\mathcal{U}}
\newcommand{\Qset}{\mathcal{Q}}
\newcommand{\Kset}{\mathcal{K}}
\newcommand{\history}{\mathcal{H}}

\newtheorem{assumption}{Assumption}[section]

\begin{document}

\title{Interpretable Knowledge Tracing via Response Influence-based Counterfactual Reasoning}

\author{
\IEEEauthorblockN{Jiajun Cui\IEEEauthorrefmark{2}~~~~Minghe Yu\IEEEauthorrefmark{3}~~~~Bo Jiang\IEEEauthorrefmark{2}~~~~Aimin Zhou\IEEEauthorrefmark{2}~~~~Jianyong Wang\IEEEauthorrefmark{4}~~~~Wei Zhang$^{*}$\IEEEauthorrefmark{2}}
\IEEEauthorblockA{\IEEEauthorrefmark{2}\textit{East China Normal University}~~~~ \IEEEauthorrefmark{3}\textit{Northeastern University}~~~~ \IEEEauthorrefmark{4}\textit{Tsinghua University} \\
cuijj96@gmail.com, yuminghe@mail.neu.edu.cn, bjiang@deit.ecnu.edu.cn\\amzhou@cs.ecnu.edu.cn, jianyong@tsinghua.edu.cn, zhangwei.thu2011@gmail.com}

\thanks{$^{*}\textrm{Corresponding author.}$ This work was supported in part by National Key R\&D Program of China (No. 2023YFC3341200), National Natural Science Foundation of China (No. 92270119, No. 62072182, No. 62137001, and No. 61977058), and Natural Science Foundation of Shanghai (No. 23ZR1418500).}}

\maketitle

\begin{abstract}

Knowledge tracing (KT) plays a crucial role in computer-aided education and intelligent tutoring systems, aiming to assess students' knowledge proficiency by predicting their future performance on new questions based on their past response records.
While existing deep learning knowledge tracing (DLKT) methods have significantly improved prediction accuracy and achieved state-of-the-art results, they often suffer from a lack of interpretability.
To address this limitation, current approaches have explored incorporating psychological influences to achieve more explainable predictions, but they tend to overlook the potential influences of historical responses.
In fact, understanding how models make predictions based on response influences can enhance the transparency and trustworthiness of the knowledge tracing process, presenting an opportunity for a new paradigm of interpretable KT.
However, measuring unobservable response influences is challenging.
In this paper, we resort to counterfactual reasoning that intervenes in each response to answer \textit{what if a student had answered a question incorrectly that he/she actually answered correctly, and vice versa}.
Based on this, we propose RCKT, a novel response influence-based counterfactual knowledge tracing framework. RCKT generates response influences by comparing prediction outcomes from factual sequences and constructed counterfactual sequences after interventions. Additionally, we introduce maximization and inference techniques to leverage accumulated influences from different past responses, further improving the model's performance and credibility.
Extensive experimental results demonstrate that our RCKT method outperforms state-of-the-art knowledge tracing methods on four datasets against six baselines, and provides credible interpretations of response influences.
The source code is available at \url{https://github.com/JJCui96/RCKT}.
\end{abstract}


\section{Introduction}\label{sec:introduction}

The rapid growth of online tutoring systems and computer-aided education has resulted in a wealth of student learning data.
Effectively harnessing this data to assess students' knowledge levels has given rise to the critical task of knowledge tracing (KT)~\cite{b_kt}.
KT involves predicting how students will perform on new questions based on their past response records. 
Accomplishing this, educators can identify learning gaps in students' knowledge and provide personalized learning materials.
Meanwhile, the advent of deep learning techniques has significantly boosted the predictive accuracy of KT.
Various approaches have emerged~\cite{b_dkt, b_lpkt, b_dimkt, b_sakt, b_akt, b_mrtkt}.
These Deep Learning Knowledge Tracing (DLKT) methods model student learning behaviors sequentially and have achieved state-of-the-art performance.

Nevertheless, a prevalent trade-off in current DLKT methods, as well as other AI and machine learning fields, is the compromise between improved performance and interpretability.
This is primarily due to the opaque nature of neural networks~\cite{black_box}.
Deep Knowledge Tracing (DKT)\cite{b_dkt} serves as a pioneer of this trend, outperforming traditional Bayesian Knowledge Tracing (BKT)\cite{b_kt}, but losing the ability to explain prediction outcomes.
Such explanations are highly valuable to educators as they provide transparency and trustworthiness.
To address the interpretability sacrifice, several researchers have developed explainable KT approaches that model student behaviors based on psychological influences.
Some machine learning-based methods~\cite{b_ktm, b_ikt} incorporate various latent traits, such as question difficulty and student ability, while achieving comparable performance with DLKT methods.
Another category of approaches focuses on integrating psychological interpretable modules into the prediction layers of their neural networks~\cite{b_tc_mirt, b_qikt}.
By adopting such methods, the resulting models can retain a higher degree of interpretability while still benefiting from the advantages of deep learning.

\begin{figure}[!t]
\centerline{\includegraphics[width=1.0\linewidth]{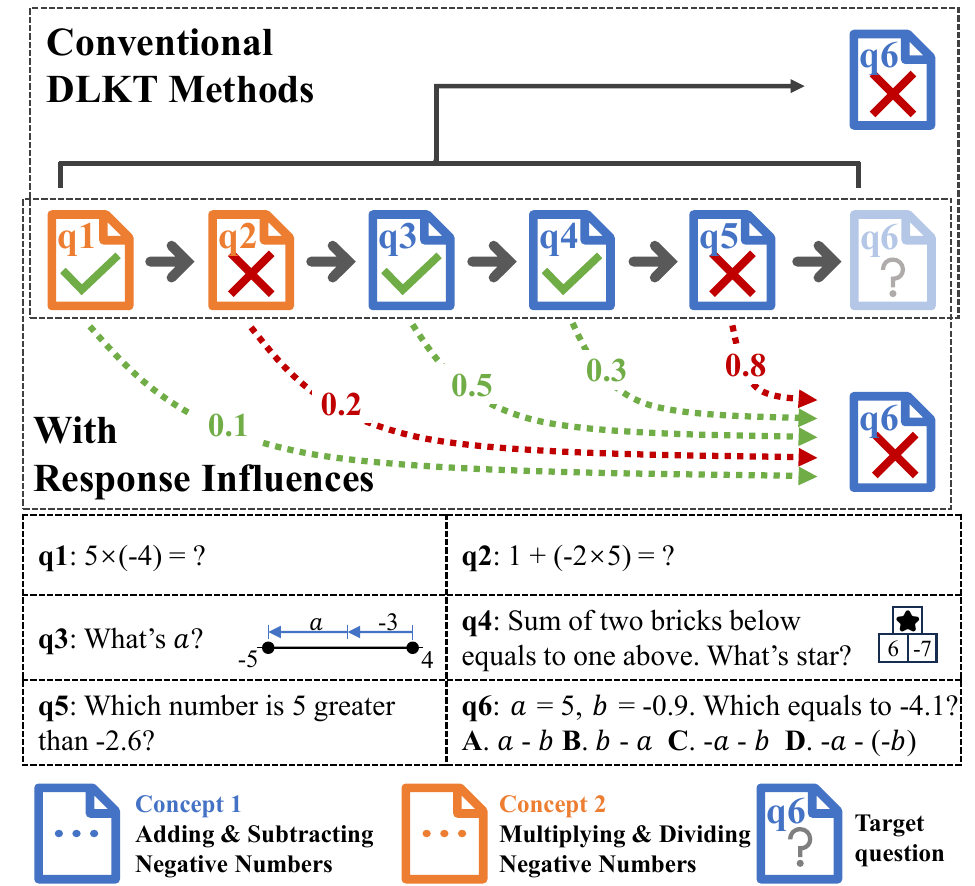}}
\vspace{-.5em}
\caption{A toy example comparing inference between conventional DLKT models (upper dashed box) and models with response influences (lower dashed box).
The arrow lines with green and red values represent the influence of past correct and incorrect responses, respectively, on the target question.
The question description is displayed in the middle.
This illustrative example is derived from the Eedi dataset, and more details about the dataset can be referred to in Sec.~\ref{subsec:experimental-setup}.}
\label{fig:intro}
\vspace{-1em}
\end{figure}

Despite the improvements in interpretability from a psychological perspective, the influences stemming from student responses remain relatively unexplored.
These influences hold valuable insights into how past responses affect students to answer new questions and grasp specific knowledge concepts.
They present an opportunity to construct more interpretable KT methods. 
By delving into these influences, educators can gain a deeper understanding of the reasons behind the formation of new responses and knowledge mastery.
This enhances the credibility of the model's prediction inference.
Fig.~\ref{fig:intro} highlights the distinctiveness of response influences.
Conventional DLKT models~\cite{b_dkt, b_lpkt, b_dimkt, b_sakt, b_akt, b_mrtkt} typically generate prediction scores by considering the entire response record as a whole.
They often stem from their internal hidden neurons, which lack sufficient interpretability~\cite{b_blackbox}.
In contrast, inference from response influences provides a clear distinction between the effects of different past responses on answering target questions.
It explicitly showcases how models make predictions.
Specifically, correct responses demonstrate a good grasp of the knowledge, thus positively impacting the ability to answer new questions correctly.
Conversely, incorrect responses indicate the opposite.
For instance, in the presented example involving the concept of \textit{adding \& subtracting negative numbers}, even though the student answered $q_3$ and $q_4$ correctly, his/her nearest wrong answer to the easy question $q_5$ leads to a larger influence in answering $q_6$ incorrectly.
Responses to $q_1$ and $q_2$ have smaller but non-zero influences since the concepts are different but relevant.
The cumulative influence of correct responses to $q_1, q_3, q_4$ is less than the one from incorrect responses to $q_2, q_5$ (0.9 vs. 1.0), resulting in the final prediction of answering the target question incorrectly.
Similar processes can be applied to determine response influences for specific knowledge concepts, as will be further explained later.
Moreover, these influences can unveil various underlying features, such as the forgetting curve and question value during student learning processes.
These insights can aid educators in improving their teaching activities, such as question recommendation and question bank construction.

Despite the potential value of response influences in KT, few studies have focused on utilizing them to enhance model interpretation.
The only similar methods employ the attention mechanism~\cite{b_transformer} to measure the similarity between responses or questions and capture their relations~\cite{b_sakt, b_akt, b_mrtkt}. 
However, these methods suffer from two main limitations when it comes to measuring response influences:
(i) The styles of multi-head and multi-layer attention mechanisms create distinct groups of similarity, which focus on different subspaces and hierarchies of sequence information~\cite{b_transformer, b_transformer_layers}.
Determining which groups imply the response influences provides no clear indication, and explaining these attention groups can be challenging.
(ii) The attention values do not necessarily reflect the true importance of the response influences, which limits their overall interpretability~\cite{b_transformer_heads, b_attention_inter}.
Additionally, the prediction outcomes are often inferred from their downstream fully-connected networks.
This leads to a lack of transparency and trustworthiness.

To address the interpretability issue in KT at the response level, we propose a novel approach called \textbf{\underline{R}esponse influence-based \underline{C}ounterfactual \underline{K}nowledge \underline{T}racing (RCKT)}.
RCKT utilizes counterfactual reasoning, a technique commonly employed in machine learning to enhance model interpretability by manipulating input samples and answering \textit{what would happen if a past event had occurred in another way}~\cite{b_cve,b_cer,b_ccr}.
For KT, our approach answers the question \textit{what if a student had answered a question incorrectly that he/she actually answered correctly, and vice versa.}
To achieve this, we create a counterfactual sequence by flipping the correctness of one past response of a student from the original factual sequence.
By comparing the predicted probability of answering the target question using both the factual and counterfactual sequences, we can effectively measure the response influence.
This allows us to gain insights into how individual responses affect the overall prediction process, which contributes to the interpretability of the KT model.
In our approach, the predicted probability is obtained through an adaptive encoder-MLP structure.
It comprises a knowledge state encoder and a multi-layer perceptron.
The encoder can adapt various sequential encoders from different DLKT methods~\cite{b_dkt, b_sakt, b_akt} to generate students' hidden knowledge states.
The MLP then combines these hidden knowledge states with the target question embedding to produce the final probability.
To generate response influences, we follow specific rules:
A decrease in the predicted probability of answering the target question correctly, resulting from flipping a correct response to be incorrect, indicates a correct response influence.
Similarly, a decrease in the probability of answering the target question incorrectly, caused by flipping an incorrect response to be correct, signifies an incorrect response influence (the definition of correct/incorrect response influence is introduced in Sec.~\ref{subsubsec:response_influence} in detail).
Then, the final prediction is made based on a comparison of the accumulated correct and incorrect response influences resulting from these flips.
Unlike traditional KT methods that employ cross-entropy loss for model training, RCKT employs a response influence-based counterfactual optimization approach.
This optimization maximizes the difference between the accumulated correct and incorrect response influences according to the ground-truth label.
Moreover, we introduce joint training for the adaptive encoder-MLP response probability generator alongside the counterfactual optimization.
This joint training approach serves to regularize the learning process, further enhancing the interpretability and overall performance of the RCKT model.

Despite the promise of our counterfactual scheme, two critical technical \textbf{challenges} remain unresolved.
Firstly, estimating unobserved counterfactual data is challenging~\cite{b_challenge}.
Flipping a response could have unpredictable effects on future responses, and also make the past responses unreliable.
To address this issue, we leverage the monotonicity assumption, a widely used theory in cognitive diagnosis~\cite{b_irt, b_mirt, b_ncd}.
This assumption posits that higher knowledge proficiency leads to a higher probability of answering questions correctly.
By employing this assumption, we prioritize preserving enough confident responses after the intervention, which provides more reliable clues to complete the counterfactual sequences.
Secondly, the process of flipping all past responses for prediction can be time-consuming.
This would lead to an increase in inference time proportionally to the sequence length.
To overcome this challenge, we propose an approximation approach, where we infer the response influences backward instead of forward.
In other words, we calculate the influence of the target question on each past response.
This approach is justified by the observation that the forward influence is positively correlated with the backward influence based on the Bayes theorem.
To calculate the correct response influences, we assume the target question is answered correctly and then flip it to be incorrect.
For the incorrect response influences, we perform the opposite procedure.
This parallelizes the inference process for measuring the influences of all responses and significantly reduces the inference time complexity by a factor of the sequence length $L$.
Under this circumstance, the knowledge state encoder is required to be bi-directional so that the model representativeness is also well improved~\cite{b_brnn,b_bert}.
By addressing these challenges, our model can achieve improved interpretability and prediction performance in KT tasks.

The main contributions are summarized as follows:
\begin{itemize}[leftmargin=*]
\item \textbf{Motivation.} We explore the response influences in KT, a crucial but often overlooked aspect in enhancing the interpretability of the DLKT methods.
To the best of our knowledge, RCKT represents the first effort that addresses the interpretation issue of DLKT at the response level.
\item \textbf{Method.} We introduce RCKT, a novel response influence-based counterfactual knowledge tracing framework. 
This approach tackles two critical technical challenges.
Firstly, it addresses the difficulty of estimating unobserved counterfactual sequences by retaining enough responses based on the monotonicity assumption.
Secondly, the time-consuming issue of inference based on flipping all past responses is resolved by approximating forward influences through backward influences.
Based on this, RCKT measures response influences via counterfactual reasoning, which effectively enhances the interpretability of the DLKT methods.
\item \textbf{Experiments.} Extensive experimental results conducted on four KT datasets demonstrate the superiority of RCKT over six baselines when combined with different DLKT knowledge state encoders.
Moreover, RCKT provides reliable interpretations of response influences.
\end{itemize}


\section{Background}\label{sec:background}

\subsection{Interpretable Knowledge Tracing}

Researchers have actively sought to address the interpretability issue of DLKT methods and have developed three main types of methods:

\subsubsection{Machine Learning}

Since the introduction of the KT task and the BKT method by Corbett and Anderson~\cite{b_kt}, machine learning-based approaches for tracking students' knowledge proficiency have become feasible.
BKT, based on the Hidden Markov Model (HMM), sequentially models and explains the student learning process by considering three aspects: the initial knowledge states, the student knowledge mastery, and the likelihood of guessing or slipping.
Subsequently, various variants of BKT~\cite{b_ktidem, b_ktforget, b_ktpps} and other machine learning methods~\cite{b_pfa, b_lfa} have emerged.
They incorporate more features and enhancements to achieve better performance.
However, the advent of DLKT methods surpassed the traditional machine learning approaches due to their superior predictive accuracy.
Nevertheless, researchers have continued to propose new machine learning-based KT methods that achieve comparable performance to DLKT methods.
For example, KTM~\cite{b_ktm} leverages a factorization machine to explore underlying student and question features, while IKT~\cite{b_ikt} employs a Tree-Augmented Naive Bayes Classifier to probe the causal relations of student ability, question difficulty, and knowledge mastery.

\subsubsection{Post-hoc Explainability} 

Post-hoc explainable methods have become popular in the field of explainable AI~\cite{b_xai}.
These methods aim to interpret neural networks after their training by deconstructing the network and attributing meanings to neurons, thus quantifying the importance of different features.
For instance, one study~\cite{b_post_hoc_kt1} binarizes learned attention weights post-training to facilitate the interpretation of learning attributes.
Another approach by Scruggs et al.\cite{b_post_hoc_kt2} involves using the means of specific internal values in DLKT models to gauge knowledge mastery.
This allows for performance comparison in post-systems.
Moreover, a previous work\cite{b_post_hoc_kt3} leverages the layer-wise relevance propagation technique to explore the relevance between input features and predictions.
Despite these advancements, existing methods still encounter challenges when it comes to interpreting the decision-making process of DLKT models.

\subsubsection{Ante-hoc Explainability}

Ante-hoc explainable methods aim to enhance model interpretability by incorporating interpretable modules directly into the model structures.
One prominent example is the attention mechanism~\cite{b_transformer}, which calculates the similarity between different instances.
Despite its popularity, DLKT methods based on the attention mechanism still lack sufficient interpretability.
Except for this, some approaches have turned to replacing neural prediction layers with psychological prediction layers.
This strategy retains the high representability of neural networks while providing explanations for prediction outcomes from a psychological perspective.
For instance, TC-MIRT~\cite{b_tc_mirt} leverages multidimensional item response theory to make final predictions, considering temporal and concept information.
Another method, QIKT~\cite{b_qikt}, predicts responses by linearly combining three explainable parameters: knowledge acquisition score, knowledge mastery score, and knowledge application score.

Our RCKT belongs to this category but adopts the response perspective instead of the psychological perspective.
This enables us to explore how past responses of students influence their answers to target questions, which has been largely overlooked in the current literature.

\begin{figure*}[!t]
\centerline{\includegraphics[width=0.95\linewidth]{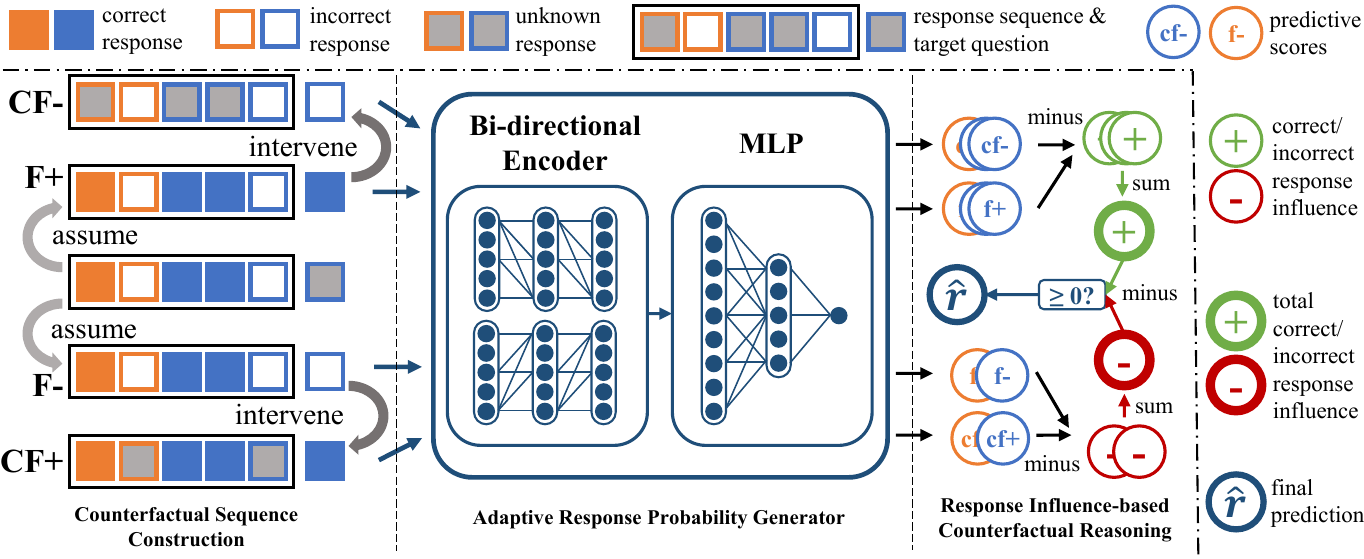}}
\caption{The framework of RCKT. $\textbf{F+}$ denotes the factual response sequence when we assume the target question is answered correctly.
$\textbf{CF-}$ denotes the counterfactual response sequence when we intervene in the response to the target question in $\textbf{F+}$ to be incorrect.
$\textbf{F-}$ and $\textbf{CF+}$ denote the sequences vice versa. The meanings of other symbols are exhibited on the top and right.}
\label{fig:RCKT}
\vspace{-1.5em}
\end{figure*}

\subsection{Counterfactual Reasoning}

The rapid development of deep learning techniques has led to the emergence of counterfactual reasoning as a new perspective to enhance the understanding of model decision processes~\cite{b_cf_exp}. 
Such methods answer \textit{what would happen if a past event had occurred in another way}, which is widely used in many AI fields.

In the domain of vision tasks, counterfactual reasoning has been employed by researchers to identify the crucial patterns that deep learning models utilize to distinguish one image category from another.
Goyal et al.~\cite{b_cve} introduced a minimum-edit problem, creating composite images from different categories to highlight their differences.
Additionally, several studies have refined this scheme by focusing on different aspects, including semantics, objects, and explanation diversity~\cite{b_scout, b_octet, b_steex, b_dive}.
In language tasks, counterfactual reasoning is commonly used to explain classification models.
For example, in a previous study~\cite{b_financial}, counterfactual explanations are generated by iteratively replacing highly relevant tokens until the prediction outcome changes.
This technique has also been applied to multi-modal tasks in various scenarios~\cite{b_cpl, b_vqa, b_navigation}.
In recommender systems, counterfactual reasoning is employed to detect the most influential historical interactions of users that impact their current preferences.
For example, in the work by~\cite{b_accent}, the impact of each item on the model's prediction is measured, and counterfactual explanations are generated.
Xu et al.~\cite{b_cr} utilized a variational auto-encoder to generate counterfactual sequences.
Furthermore, some studies combine counterfactual reasoning with other perspectives, such as review-based recommendation and logical reasoning~\cite{b_cer, b_ccr}. 

Counterfactual reasoning has demonstrated its value in various fields, but its potential in DLKT remains largely untapped.
Most of the existing explainable methods in DLKT focus on interpretability from psychological perspectives, often overlooking the importance of input responses~\cite{b_tc_mirt, b_qikt}.
We propose RCKT to fill this gap, and in a more fine-grained way.
In other words, in contrast to those mainstream counterfactual reasoning methods that minimally edit input samples to change the outcome, RCKT quantifies the influence of each response on the model's prediction. By doing so, the entire model decision process becomes transparent and concrete.

Moreover, counterfactual reasoning is also leveraged for improving model unbiasedness or fairness~\cite{b_cf_other}.
In this paper, we only focus on enhancing model interpretability.
It is also worth noting that Influence Functions~\cite{b_if} follow a similar form by quantifying the significance of removing or modifying individual training samples.
This method uses a gradient-based approach to estimate the influences of some training samples contributing to the entire trained model, aiming to enhance model robustness and generalization.~\cite{b_if_inter, b_if_sharp}.
In contrast, methods based on counterfactual reasoning, such as our RCKT, determine the influences of manipulating specific parts of an input sample on its predictive output to explain the output from the input.

\section{Preliminary}

\subsection{Deep Learning Knowledge Tracing}

The KT task's primary objective is to predict students' proficiency in various knowledge concepts by assessing their performance on newly presented questions.
Formally, we are given a student set $\Uset$, a question set $\Qset$, and a knowledge concept set $\Kset$.
The aim is to determine if a student $u\in\Uset$ can correctly answer a target question $q^u_{t+1}\in\Qset$, given its corresponding knowledge concept set $\Kset_{t+1}^u \subset \Kset$.
This prediction is based on the student's historical response sequence up to time $t$, denoted as $\history^u_t=\{(q^u_1,r^u_1,\Kset^u_1),(q^u_2,r^u_2,\Kset^u_2),...,(q^u_t,r^u_t,\Kset^u_t)\}$, where $r^u_t \in \{0,1\}$ indicates the binary correctness of the response.
Throughout the rest of this paper, we will omit the superscript of individual users unless specified otherwise.
Since knowledge proficiency is not directly observable, such response correctness prediction conventionally acts as a proxy to measure the quality of knowledge tracing~\cite{b_kt_survey}.
Consequently, DLKT models have emerged, capitalizing on the power of neural networks to achieve state-of-the-art performance.
By this means, the estimated binary correctness is obtained by
\begin{equation}
\label{eq:kt}
\hat{r}_{t+1}=\mathbbm{1}\left(h\left(q_{t+1},\Kset_{t+1},\history_t|\boldsymbol{\Theta}\right)\geq\gamma\right),
\end{equation}
where $\mathbbm{1}(\cdot)$ denotes the indicator function and $\gamma$ represents the threshold that discriminates the output continuous values (i.e., predictive scores) into binary ones.
The function $h(\cdot)$ indicates the deep neural network and $\boldsymbol{\Theta}$ denotes its abundant but hard-to-explain learnable parameters.

\subsection{Ante-hoc Interpretable Deep Learning Knowledge Tracing}

Ante-hoc explainability requires the model to embed interpretable modules into the network structure.
As previously mentioned, the attention mechanism is one approach that falls under this category but may not offer sufficient interpretability. 
Therefore, we formulate the ante-hoc interpretable DLKT according to the common approaches~\cite{b_tc_mirt,b_qikt} that incorporate interpretable modules into the prediction layers as follows:
\begin{equation}
\label{eq:exp_kt}
\hat{r}_{t+1}=\mathbbm{1}\left(g\left(\textbf{f}\left(q_{t+1},\Kset_{t+1},\history_t|\boldsymbol{\Phi}\right)|\boldsymbol{\Omega}\right)\geq\gamma\right).
\end{equation}
Similarly, $\gamma$ is the decision threshold.
$\boldsymbol{\Phi}$ is the learnable parameters of the upstream neural network $\textbf{f}(\cdot)$ that outputs a set of interpretable values.
$g(\cdot)$ is the interpretable module that generates predictive scores by the interpretable parameters $\boldsymbol{\Omega}$.
These interpretable parameters could be learnable or pre-set, which clearly reflect the model decision process.

\subsection{Monotonicity Assumption}

Monotonicity Assumption is a common principle in cognitive diagnosis~\cite{b_irt,b_mirt,b_ncd} that states:\\
\begin{assumption}
\textit{The higher the proficiency of a student in any knowledge concept, the more likely the student can answer a question correctly.\\}
\end{assumption}
This assumption establishes the inherent connection between knowledge proficiency and student responses, forming the foundation for evaluating KT models.
The accuracy of predicting student responses serves as a crucial indicator of the knowledge tracing quality. 

We leverage this assumption for retaining reliable response sequences after the counterfactual intervention.
The details of this process will be elaborated later.

\section{Method}\label{sec:method}

\begin{figure}[!t]
\centerline{\includegraphics[width=0.8\linewidth]{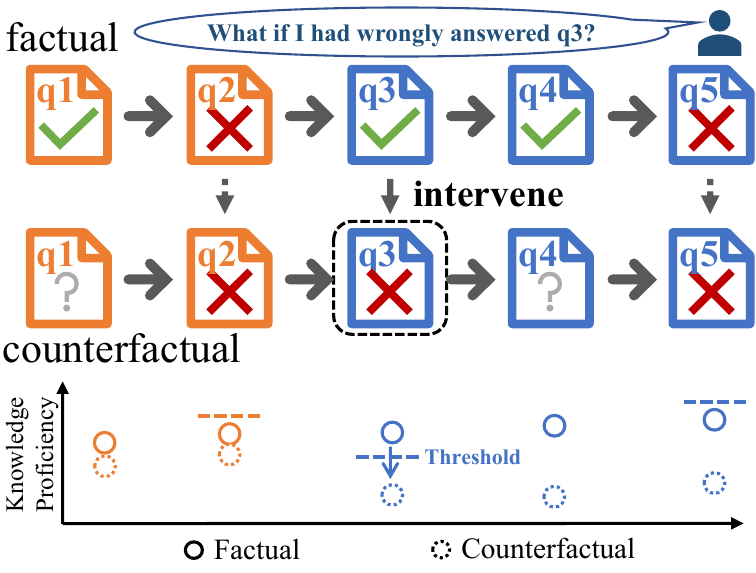}}
\caption{The process of constructing counterfactual sequences by the monotonicity assumption. We use the sample example in Fig.~\ref{fig:intro}. Thresholds determine whether the knowledge proficiency is enough to make a correct answer.}
\label{fig:cf}
\vspace{-1.5em}
\end{figure}

\subsection{Overview}

RCKT utilizes response influence-based counterfactual reasoning to create an interpretable knowledge tracing framework centered around student responses. The methodology comprises three main procedures:
Firstly, counterfactual sequence construction involves retaining reliable responses after intervening in each response. This retention is guided by the monotonicity assumption.
Secondly, an adaptive response probability generator takes both the original factual and constructed counterfactual sequences as input, and generates the probabilities of correctly or incorrectly answering questions.
Finally, response influence-based counterfactual reasoning calculates the influence of each response by contrasting the generated probabilities from the factual and counterfactual sequences.
The accumulations of influences from both correct and incorrect responses are then compared to conduct a response influence-based counterfactual optimization and lead to the final prediction.
To address time-consumption issues, a response influence approximation is proposed, which reverses the influence direction.

The overall framework of RCKT after the response influence approximation, is illustrated in Fig.~\ref{fig:RCKT}.
For a logical introduction to RCKT, we explain the counterfactual sequence construction at first, then the response influence-based counterfactual reasoning, and then the adaptive response probability generator in what follows.


\subsection{Counterfactual Sequence Construction}

To estimate the influence of each historical response on the target question, we perform an intervention by flipping its correctness, and then observe how it impacts the prediction outcome. However, directly flipping a response may adversely affect the reliability of other responses.
This is because the student's knowledge state undergoes changes due to this intervention.
To mitigate this issue, we employ the monotonicity assumption, which allows us to maintain the reliability of response sequences through two key operations: \textbf{masking} and \textbf{retaining}. 

Take the same example as illustrated in Fig.~\ref{fig:cf}. Based on the monotonicity assumption, when we intervene in the student's response to $q_3$ by changing it from correct to incorrect, it indicates a decrease in the student's proficiency in the concept of \textit{adding \& subtracting negative numbers}.
This decrease in proficiency has two primary consequences: (i) It will persist and potentially affect the subsequent responses to $q_4$ and $q_5$. (ii) It may render the prior responses to $q_1$ and $q_2$ unreliable as evidence to support the incorrect response to $q_3$.
Considering the first circumstance, we cannot determine definitively whether the proficiency decrease would result in the correct response to $q_4$ becoming incorrect. Therefore, we \textbf{mask} the correctness of this response in the counterfactual world as unknown.
For $q_5$, however, we can confidently infer that the decrease in proficiency would not impact its incorrect response. In other words, both the proficiency before and after the change are below the threshold, and therefore, we \textbf{retain} this response as incorrect in the counterfactual world.
A similar process is applied to the prior responses to $q_1$ and $q_2$, even though they involve a different knowledge concept \textit{multiplying \& dividing}. The logic of masking and retaining still applies due to the underlying relations between these concepts.
Likewise, the process of flipping an incorrect response to be correct is also applicable.
In this case, the knowledge proficiency increases, so we retain the correct responses and mask the incorrect ones instead.

We omit the knowledge concepts for conciseness to formulate this process.
Suppose there is a response sequence $\history_t=\{(q_1,r_1),...,(q_t,r_t)\}$.
We use $(Q,R)_{1:t}$ to denote the ordered set consisting of the $1^{st}$ to $t^{th}$ responses.
The response sequence could be then represented in the factual world as
\begin{equation}\label{eq:factal_set}
\begin{aligned}
F_t&=\{(Q,R)_{1:t}\}\\
&=\{(Q,R)^+_{1:t},(Q,R)^-_{1:t}\},
\end{aligned}  
\end{equation}
where $(Q,R)^+_{1:t}$ and $(Q,R)^-_{1:t}$ denote the set of correct and incorrect responses, respectively.
When flipping a correct response $(q_i, r_i)^+$, a counterfactual response sequence is generated. Utilizing the monotonicity assumption, we retain the incorrect responses and mask the correct ones to create the counterfactual sequence.
It is denoted as:
\begin{equation}\label{eq:counterfactal_neg_set}
CF_{t,i^-}=\{(q_i,r_i)^-,(Q,R)^-_{1:t/\{i\}},(Q,M)^+_{1:t/\{i\}}\}, 
\end{equation}
where we replace $R$ with $M$ to signify the masking operation.
The subscript $1:t/\{i\}$ denotes the index from 1 to $t$ except for $i$.
Conversely, if the response is initially incorrect, the counterfactual sequence would be reversed as:
\begin{equation}\label{eq:counterfactal_pos_set}
CF_{t,i^+}=\{(q_i,r_i)^+,(Q,R)^+_{1:t/\{i\}}, (Q,M)^-_{1:t/\{i\}}\}.
\end{equation}
By creating these counterfactual sequences, we can measure the influence of the $i^{th}$ response by comparing the prediction outcomes from the factual and counterfactual sequences related to the target question.
To make such a comparison easy to understand, we rewrite the factual set in Eq.~\ref{eq:factal_set} from the view of a single response and make it symmetry with Eq.~\ref{eq:counterfactal_neg_set} and~\ref{eq:counterfactal_pos_set} as follows:
\begin{equation}\label{eq:factal_set_rewrite}
\begin{aligned}
F_{t,i^+}&=\{(q_i,r_i)^+,(Q,R)^+_{1:t/\{i\}}, (Q,R)^-_{1:t/\{i\}}\}\\
F_{t,i^-}&=\{(q_i,r_i)^-,(Q,R)^+_{1:t/\{i\}}, (Q,R)^-_{1:t/\{i\}}\}.
\end{aligned}  
\end{equation}
$F_{t,i^+}$ denotes the factual set when the $i^{th}$ response is correct and $F_{t,i^-}$, vice versa.

\subsection{Response Influence-based Counterfactual Reasoning}\label{subsec:ricr}
Counterfactual reasoning refers to editing model input to alternate the outcome.
This paper leverages this scheme in a more fine-grained way to measure each response's influence on answering the target question.

\subsubsection{Response Influence Measurement}\label{subsubsec:response_influence}
To measure response influences, we compare the probabilities of answering questions correctly or incorrectly in the factual and counterfactual worlds.
For a correct response, we flip it to be incorrect and observe the resulting drop in the probability of answering the target question correctly. We define this probability drop as a \textbf{correct response influence}.
For a correct response $(q_i,r_i)^+$, given the factual sequence $F_{t,i^+}$, the probability to answer the target question $q_{t+1}$ correctly is
\begin{equation}\label{eq:f_pos_score}
f_{i^+\to (t+1)^+} = p(r_{t+1}=1|q_{t+1},F_{t,i^+}).
\end{equation}
After the intervention, the response is flipped to be incorrect, and the new probability in the counterfactual world becomes:
\begin{equation}\label{eq:cf_neg_score}
cf_{i^-\to (t+1)^+} = p(r_{t+1}=1|q_{t+1},CF_{t,i^-}).
\end{equation}
We can then calculate the correct response influence by subtracting the counterfactual probability from the factual probability:
\begin{equation}\label{eq:delta_pos}
\Delta_{i^+\to (t+1)^+}=f_{i^+\to (t+1)^+}-cf_{i^-\to (t+1)^+}
\end{equation}
subject to
\begin{equation}\label{eq:delta_st}
\Delta_{i^+\to (t+1)^+}\geq0.
\end{equation}
This constraint ensures that the probability of answering the target question correctly should decrease when we intervene in a correct response to be incorrect. This decrease in probability reflects the reduction in knowledge proficiency due to the intervention.

Similarly, we derive an \textbf{incorrect response influence}, by flipping an incorrect response to be correct and obtaining the drop in the probability of answering the target question incorrectly:
\begin{equation}\label{eq:delta_neg}
\begin{aligned}
&f_{i^-\to (t+1)^-} = p(r_{t+1}=0|q_{t+1},F_{t,i^-}),\\
&cf_{i^+\to (t+1)^-} = p(r_{t+1}=0|q_{t+1},CF_{t,i^+}),\\
&\Delta_{i^-\to (t+1)^-}=f_{i^-\to (t+1)^-}-cf_{i^+\to (t+1)^-}\\
&\quad\quad\quad\text{s.t., }\Delta_{i^-\to (t+1)^-} \geq0.
\end{aligned}  
\end{equation}
Afterwards, we get the correct and incorrect response influences for further reasoning.

\subsubsection{Response Influence-based Counterfactual Reasoning}

We present an interpretable predictive framework that integrates the correct and incorrect response influences to derive the final prediction.
Specifically, the total correct and incorrect response influences of these responses are first computed by
\begin{equation}\label{eq:total_influence}
\Delta^+_{t+1}=\sum_{i^+}^t\Delta_{i^+\to(t+1)^+},\quad\Delta^-_{t+1}=\sum_{i^-}^t\Delta_{i^-\to(t+1)^-},
\end{equation}
where $i^+$ and $i^-$ indicate the indices of the correct and incorrect responses, respectively.
We then make the final prediction by comparing the total correct and incorrect response influences of the responses on answering the target question.
A correct response to the target question indicates a larger total correct response influence than the total incorrect response influence, and vice versa.
This is expressed as
\begin{equation}\label{eq:rckt_pred}
\hat{r}_{t+1}=\mathbbm{1}\left(\Delta^+_{t+1}-\Delta^-_{t+1}\geq0\right).
\end{equation}

This reasoning scheme is transparent and explainable because it simply sums up the different influences of the responses on answering the target question.
The final prediction is made by comparing the total correct and incorrect response influences, which is a reasonable and easy-to-understand process.
It is worth noting that this framework falls into the ante-hoc explainability category and also follows Eq.~\ref{eq:exp_kt}.
In our case, the interpretable function $g(\cdot|\boldsymbol{\Omega})$ represents the addition and subtraction of the response influences.
The threshold $\gamma$ is set to zero.
Besides, all the referred probability during this counterfactual reasoning is generated by an adaptive encoder-MLP neural network, which will be introduced later. 

\subsubsection{Response Influence-based Counterfactual Optimization}
In contrast to traditional KT methods that use binary cross-entropy (BCE) loss to optimize models based on predicted scores and true labels, we adopt a response influence-based counterfactual optimization scheme.
It maximizes the difference between total correct and incorrect response influences during counterfactual reasoning.
To achieve this, we formulate the optimization based on the condition mentioned earlier, where answering the target question correctly requires the total correct response influence to be larger than the total incorrect response influence, and vice versa:
\begin{equation}\label{eq:optimize_compare}
\begin{aligned}
&\text{if }r_{t+1}=1\textbf{},\quad\Delta^+_{t+1}\geq\Delta^-_{t+1},\\
&\text{if }r_{t+1}=0,\quad\Delta^+_{t+1}\leq\Delta^-_{t+1}.
\end{aligned}  
\end{equation}
With this condition, we perform the optimization as
\begin{equation}\label{eq:optimize}
\begin{aligned}
&\max\left(\Delta^-_{t+1}-\Delta^+_{t+1}\right)\cdot(-1)^{r_{t+1}}\\
&\quad\quad\text{s.t., }\Delta_{i^+\to (t+1)^+}\geq0,\,\,\Delta_{i^-\to (t+1)^-}\geq0.
\end{aligned}
\end{equation}
To convert this optimization into a loss function to minimize, we use a negative logarithmic form as follows:
\begin{equation}\label{eq:cf_loss}
    \mathcal{L}_{CF}=-\log\left(\frac{(-1)^{r_{t+1}}}{2t}\left(\Delta^-_{t+1}-\Delta^+_{t+1}\right)+\frac{1}{2}\right) + \alpha\mathcal{L}^*
\end{equation}
with a constraint term
\begin{equation}\label{eq:cf_cs}
    \mathcal{L}^*\!=\!\sum_{i^+}^t\max\left(-\Delta_{i^+\to (t+1)^+},\!0\right)\!+\!\sum_{i^-}^t\max\left(-\Delta_{i^-\to (t+1)^-},\!0\right)
\end{equation}
to regularize each response influence to be not smaller than zero.
The $2t$ in the denominator and the term $1/2$ aim to scale the difference into $(0,1)$ for the logarithmic function.
The operator $\max$ here indicates the function taking the maximum value from input instead of the one making maximization in Eq.~\ref{eq:optimize}.
$\alpha$ is a hyper-parameter to control the intensity of the constraints, which we set to 1.0 in practice.
We employ the negative logarithmic form instead of direct optimization.
This is because it applies larger punishment when the difference between the total correct and incorrect response influences is close to zero.
This intensifies the counterfactual optimization process, which effectively encourages the two response influences to diverge from each other.

\subsubsection{Response Influence Approximation}\label{subsec:ria}

In practice, getting the probability of answering the target question correctly or incorrectly requires processing the given response sequence by a sequential model.
Most KT methods use this paradigm to predict student performance.
For instance, considering DKT as an example, its Long Short-Term Memory (LSTM) sequential network processes a single $t$-length response sequence with a time complexity of $O(td^2)$, where $d$ is the number of hidden dimensions. However, calculating the response influences in RCKT necessitates generating $t$ different counterfactual response sequences, leading to a $t$-fold increase in inference time.
This would result in a time complexity of $O(t^2d^2)$.
We address this issue by introducing an approximation for the response influences.
Specifically, instead of generating multiple counterfactual response sequences, we use the influences on the past responses from intervening in assumed responses to the target question (backward), to estimate the influences on the target question from the past responses (forward), i.e.,
\begin{equation}\label{eq:approx}
\begin{aligned}
&\Delta_{(t+1)^+\to i^+}\sim\Delta_{i^+\to (t+1)^+},\\
&\Delta_{(t+1)^-\to i^-}\sim\Delta_{i^-\to (t+1)^-}.
\end{aligned}
\end{equation}
Here, $\Delta_{(t+1)^+\to i^+}$ represents the change in predicting a correct historical response when we assume the target question is correct and intervene it to be incorrect, and vice versa.
This operation is visualized on the left side of Fig.~\ref{fig:RCKT}, which also considers the example shown in Fig.\ref{fig:intro} as graphic symbols.
In this example, when the target question changes from being answered correctly to being answered incorrectly, we retain the two incorrect responses and mask the three correct ones. Conversely, when the target question is assumed to be answered incorrectly, the opposite holds true.
We thus rewrite Eq.~\ref{eq:counterfactal_neg_set},~\ref{eq:factal_set_rewrite},~\ref{eq:f_pos_score},~\ref{eq:cf_neg_score} and~\ref{eq:delta_pos} as

\begin{equation}\label{eq:approx_rewrite}
\begin{aligned}
&F_{t/i,(t\!+\!1)^+}\!=\!\{(q_{t\!+\!1},r_{t\!+\!1})^+,(Q,R)^+_{1:t/\{i\}}, (Q,R)^-_{1:t/\{i\}}\},\\
&CF_{t/i,(t\!+\!1)^-}\!=\!\{(q_{t\!+\!1},r_{t\!+\!1})^-,(Q,R)^-_{1:t/\{i\}}, (Q,M)^+_{1:t/\{i\}}\},\\
&\quad\quad\quad f_{(t+1)^+\to i^+} = p(r_{i}=1|q_{i},F_{t/i,(t+1)^+}),\\
&\quad\quad\quad cf_{(t+1)^-\to i^+} = p(r_{i}=1|q_{i},CF_{t/i,(t+1)^-}),\\
&\quad\quad\quad \Delta_{(t+1)^+\to i^+}=f_{(t+1)^+\to i^+}-cf_{(t+1)^-\to i^+}.
\end{aligned}
\end{equation}
Similarly, there is
\begin{equation}\label{eq:approx_neg}
\Delta_{(t+1)^-\to i^-}=f_{(t+1)^-\to i^-}-cf_{(t+1)^+\to i^-}.
\end{equation}
We use the Bayes formula to infer this approximation from
\begin{equation*}\label{eq:bayes}
\frac{f_{(t\!+\!1)^+\!\to i^+}}{f_{i^+\to (t\!+\!1)^+}}\!=\!\frac{p(r_{i}\!=\!1|q_i,\!q_{t\!+\!1},\!(Q,\!R)^+_{1:t/\{i\}},\!(Q,\!R)^-_{1:t/\{i\}})}{p(r_{t\!+\!1}\!=\!1|q_i,\!q_{t\!+\!1},\!(Q,\!R)^+_{1:t/\{i\}},\!(Q,\!R)^-_{1:t/\{i\}})},
\end{equation*}
and
\begin{equation}
\begin{aligned}
&\frac{1\!-\!cf_{(t\!+\!1)^-\!\to i^+}}{1\!-\!cf_{i^-\to (t\!+\!1)^+}}=\\
&\quad\quad\frac{1\!-\!p(r_{i}\!=\!1|q_i,\!q_{t\!+\!1},\!(Q,\!M)^+_{1:t/\{i\}},\!(Q,\!R)^-_{1:t/\{i\}})}{1\!-\!p(r_{t\!+\!1}\!=\!1|q_i,\!q_{t\!+\!1},\!(Q,\!M)^+_{1:t/\{i\}},\!(Q,\!R)^-_{1:t/\{i\}})}.
\end{aligned}
\end{equation}
The right-hand sides of the equal signs show the individual probability of answering the target or historical question correctly under the same sequence conditions.
Therefore, the backward and forward influences involved in the left part are positively correlated.
Then we have
\begin{equation}\label{eq:delta_sum_rev}
\Delta_{t+1}^+=\sum_{i^+}^t\Delta_{(t+1)^+\to i^+},\quad\Delta_{t+1}^-=\sum_{i^-}^t\Delta_{(t+1)^-\to i^-}.
\end{equation}
As depicted on the right side of Fig.~\ref{fig:RCKT}, this example generates three correct and two incorrect response influences, and their sums are compared to make the final prediction.
By this approximation, all the counterfactual interventions are applied to the target question, which means we only need to construct two counterfactual response sequences instead of $t$ as in the original method.
This leads to a significant reduction in inference time compared to processing the original $t$ sequences separately.
Taking the same example of applying LSTM, the time complexity decreases to $O(td^2)$ as the same as DKT.
The response influence-based counterfactual reasoning after approximation is presented on the right side of Fig.~\ref{fig:RCKT}.

Tab.~\ref{tab:run} illustrates the process of RCKT inferring the same example after response influence approximation. As shown, $q_6$ is assumed to be answered correctly and is then flipped to an incorrect response. The probabilities of correctly answering the questions with correct responses ($q_1, q_3$, and $q_4$) before and after this flip are then compared to generate the correct response influences. Conversely, we can derive the influences for the two incorrect responses. Therefore, the total correct and incorrect response influences are compared to make the final prediction (0.9 vs. 1.0).

\begin{table}[!t]
\setlength{\tabcolsep}{1.5pt}
\begin{center}
\caption{An example showing how RCKT uses response influence approximation to conduct counterfactual reasoning. ``$\circ$'' denotes the mask operation.}
\begin{tabular}{lcccccc|lcccccc}
\hline\hline
\multicolumn{7}{c|}{Assuming $r_6=1$}                                                                                                                 & \multicolumn{7}{c}{Assuming $r_6=0$}                                                                                                                  \\ \hline
\multicolumn{1}{l|}{Question}                  & $q_1$        & $q_2$    & $q_3$        & $q_4$        & \multicolumn{1}{c|}{$q_5$}    & $q_6$        & \multicolumn{1}{l|}{Question}                  & $q_1$        & $q_2$    & $q_3$        & $q_4$        & \multicolumn{1}{c|}{$q_5$}    & $q_6$        \\ \hline
\multicolumn{1}{l|}{$F_{t,(t+1)^+}$}           & $\checkmark$ & $\times$ & $\checkmark$ & $\checkmark$ & \multicolumn{1}{c|}{$\times$} & $\checkmark$ & \multicolumn{1}{l|}{$F_{t,(t+1)^-}$}           & $\checkmark$ & $\times$ & $\checkmark$ & $\checkmark$ & \multicolumn{1}{c|}{$\times$} & $\times$     \\
\multicolumn{1}{l|}{$f_{(t+1)^+\to i^+}$}      & 0.6          & -        & 0.7          & 0.6          & \multicolumn{1}{c|}{-}        & -            & \multicolumn{1}{l|}{$f_{(t+1)^-\to i^-}$}      & -            & 0.6      & -            & -            & \multicolumn{1}{c|}{0.9}      & -            \\
\multicolumn{1}{l|}{$CF_{t,(t+1)^-}$}          & $\circ$            & $\times$ & $\circ$            & $\circ$            & \multicolumn{1}{c|}{$\times$} & $\times$     & \multicolumn{1}{l|}{$CF_{t,(t+1)^+}$}          & $\checkmark$ & $\circ$        & $\checkmark$ & $\checkmark$ & \multicolumn{1}{c|}{$\circ$}        & $\checkmark$ \\
\multicolumn{1}{l|}{$cf_{(t+1)^-\to i^+}$}     & 0.5          & -        & 0.2          & 0.3          & \multicolumn{1}{c|}{-}        & -            & \multicolumn{1}{l|}{$cf_{(t+1)^+\to i^-}$}     & -            & 0.4      & -            & -            & \multicolumn{1}{c|}{0.1}      & -            \\ \hline
\multicolumn{1}{l|}{$\Delta_{(t+1)^+\to i^+}$} & 0.1          & -        & 0.5          & 0.3          & \multicolumn{1}{c|}{-}        & -            & \multicolumn{1}{l|}{$\Delta_{(t+1)^-\to i^-}$} & -            & 0.2      & -            & -            & \multicolumn{1}{c|}{0.8}      & -            \\ \hline
\multicolumn{7}{c|}{$\Delta_{t+1}^+=0.1+0.5+0.3=0.9$}                                                                                                             & \multicolumn{7}{c}{$\Delta_{t+1}^-=0.2+0.8=1.0$}                                                                                                              \\ \hline\hline
\end{tabular}
\label{tab:run}
\end{center}
\vspace{-2em}
\end{table}

\subsection{Adaptive Response Probability Generator}
During the counterfactual reasoning, we leverage the power of neural networks to generate the accurate probability of answering given questions correctly, i.e., estimating the probability in Eq.~\ref{eq:approx_rewrite}.
\subsubsection{Adaptive Network Structure}
The adaptive encoder-MLP structure in RCKT consists of a sequential knowledge state encoder and a predictive MLP.
Before encoding, each input question $q_i$ is embedded with its related knowledge concepts $\Kset_i$ using
\begin{equation}\label{eq:question_emb}
\textbf{e}_i=\textbf{q}_i+\frac{1}{|\Kset_i|}\sum_{k_j\in\Kset_i}\textbf{k}_j,
\end{equation}
where $\textbf{q}_i\in\mathbb{R}^{1\times d}$ and $\textbf{k}_j\in\mathbb{R}^{1\times d}$ denote the ID embeddings of $q_i$ and concept $k_j$, respectively.
$d$ is the size of the hidden dimension in RCKT.
Consequently, we fuse the binary correctness by three categories: the response is incorrect, correct, or masked in the reasoning.
These categories are denoted by $\tilde{r}_i\in\{0,1,2\}$, and $\textbf{r}_i\in\mathbb{R}^{1\times d}$ is the corresponding category embedding.
Then the response embedding is obtained by
\begin{equation}\label{eq:response_emb}
\textbf{a}_i=\textbf{e}_i+\textbf{r}_i.
\end{equation}
To fulfill the response influence approximation requirements, the encoder is bi-directional.
It predicts intermediate responses given past and future responses as presented in the middle part of Fig.~\ref{fig:RCKT}.
Suppose $\textbf{A}_{1:t}$ denotes the set of response embeddings from $\textbf{a}_1$ to $\textbf{a}_t$.
The hidden knowledge state is derived by
\begin{equation}\label{eq:encoder}
\textbf{h}_i =\overrightarrow{\textbf{Enc}}(\textbf{A}_{1:i-1}) + \overleftarrow{\textbf{Enc}}(\textbf{A}_{i+1:t+1}),
\end{equation}
where $\overrightarrow{\textbf{Enc}}(\cdot)$ and $\overleftarrow{\textbf{Enc}}(\cdot)$ represent the two directions of the encoder.
When $i=1$ (i.e., the first response), we directly use $\overleftarrow{\textbf{Enc}}(\textbf{A}_{2:t+1})$ as the output.
Afterwards, we generate the probability of answering $q_i$ correctly by an MLP with its question embedding:
\begin{equation}\label{eq:mlp}
p_i=\sigma\left(\text{ReLU}\left([\textbf{h}_i\oplus\textbf{e}_i]\textbf{W}_1+\textbf{b}_1\right)\textbf{W}_2+\textbf{b}_2\right).
\end{equation}
Here, $\textbf{W}_1\in\mathbb{R}^{2d\times d},\textbf{W}_2\in\mathbb{R}^{d\times1},\textbf{b}_1\in\mathbb{R}^{1\times d}$ and $\textbf{b}_2\in\mathbb{R}^{1\times 1}$ represent the learnable parameters in the MLP. $\oplus$ denotes the concatenation operation and $\sigma(\cdot)$ denotes the sigmoid function.
$\text{ReLU}(\cdot)$ is the activation function.
Accordingly, the probability of answering the question incorrectly is $1-p_i$.
It's important to note that this encoder can be adapted to multiple KT sequence encoders, as they can be extended bi-directionally.
In practice, the sequence encoders from DKT, SAKT, and AKT are used and extended in a multi-layer style, and we make them stackable in the RCKT framework.

\subsubsection{Joint Training}

To ensure robust probability estimation for calculating response influences, the adaptive response probability generator is jointly trained along with the response influence-based counterfactual optimization.
The probability generation from factual sequences is directly trained using the original sequences. For a question $q_i$, given its contextual response sequence $\{(Q,R)^+_{1:t/\{i\}}, (Q,R)^-_{1:t/\{i\}}\}$, the probability of answering $q_i$ correctly is derived from the adaptive response probability generator and denoted as $p_i^F$. The standard BCE loss with the ground-truth correctness $r_i$ is then used for training this probability generation:
\begin{equation}\label{eq:loss
_factual}
\mathcal{L}_{F}=-\frac{1}{t}\sum_i^tr_i\log(p_i^{F})+(1-r_i)\log(1 - p_i^{F}).
\end{equation}
For the probability from the counterfactual sequences, however, the actual response records do not provide sequences in the counterfactual world.
Inspired by contrastive learning of sequential modeling~\cite{b_cf}, we address this issue by augmenting the sequences from masking responses as unknown.
For a question $q_i$, given its contextual response sequence with the incorrect responses masked, $\{(Q,R)^+_{1:t/\{i\}}, (Q,M)^-_{1:t/\{i\}}\}$, the probability of answering $q_i$ correctly is then derived and denoted as $p_i^{M+}$.
Similarly, we obtain $p_i^{M-}$ by given $\{(Q,R)^-_{1:t/\{i\}}, (Q,M)^+_{1:t/\{i\}}\}$, the contextual response sequence with the correct responses masked.
The masking scheme to augment response sequences has two advantages: (i) It creates similar conditions for generating probabilities from counterfactual sequences that contain all correct or incorrect responses with others masked. (ii) Masking responses as unknown does not undermine the original sequences' authenticity.
For the probability $p_i^{M^+}$ and $p_i^{M-}$, we also have
\begin{equation}\label{eq:loss
_counterfactuaal}
\begin{aligned}
&\mathcal{L}_{M^+}=-\frac{1}{t}\sum_i^tr_i\log(p_i^{M^+})+(1-r_i)\log(1 - p_i^{M^+}),\\
&\mathcal{L}_{M^-}=-\frac{1}{t}\sum_i^tr_i\log(p_i^{M^-})+(1-r_i)\log(1 - p_i^{M^-}).\\
\end{aligned}
\end{equation}
Finally, we jointly optimize all the four losses as
\begin{equation}\label{eq:final_loss}
    \mathcal{L} = \mathcal{L}_{CF} + \lambda\left(\mathcal{L}_{F} + \mathcal{L}_{M^+} + \mathcal{L}_{M^-}\right),
\end{equation}
where $\lambda$ is a hyper-parameter for balance.
It is worth noting that this total loss is for one response sequence with one target question, i.e., one training sample in the dataset.
In practice, we use the average loss over all student responses to train RCKT.
For simplicity, we leave out the averaging notations.
To alleviate the overfitting problem, we also apply the dropout technique in the middle of MLP and add $l_2$ normalization in the loss function, which are omitted in our formulation.  

\begin{table}[t]
\begin{center}
\caption{Statistics of the four preprocessed datasets.}
\begin{tabular}{l|ccc}
\hline
\hline
Dataset            & ASSIST09 & ASSIST12 & Slepemapy  \\ \hline
year of collection    & 09-10    & 12-13    & 13-15  \\
\#response         & 0.4m     & 2.7m     & 10.0m  \\
\#sequence         & 10.7k    & 62.6k    & 234.5k \\
\#question         & 13.5k    & 53.1k    & 2.2k  \\
\#concept          & 151      & 265      & 1458    \\
\#concept/question & 1.22     & 1        & 1    \\ 
\%correct responses & 0.63 & 0.70 & 0.78\\
\hline\hline
\end{tabular}
\label{tab:stat}
\end{center}
\end{table}

\begin{table}[!t]
\setlength{\tabcolsep}{1.5pt}
\begin{center}
\caption{Hyper-parameter setting of RCKT.}
\vspace{-1.5em}

\begin{tabular}{l|ccc}
\hline\hline
Encoder  & DKT                 & SAKT                & AKT                \\ \hline
ASSIST09  & \{1e-3,0.1,1e-5,0.3,2\} & \{2e-3,0.1,2e-4,0.2,3\} & \{5e-4,0.01,5e-5,0,3\} \\
ASSIST12  & \{2e-3,0.01,1e-5,0,3\}  & \{2e-3,0.1,5e-4,0.2,3\} & \{5e-4,0.05,1e-5,0,3\} \\
Slepemapy & \{1e-3,0.1,0,0,3\}      & \{5e-4,0.4,1e-5,0,3\}   & \{5e-4,0.01,1e-5,0,2\} \\
Eedi      &      \{1e-3,0.1,0,0,3\}  &     \{1e-3,0.1,1e-5,0,3\}        &             \{5e-4,0.01,1e-5,0,3\}                    \\ \hline\hline
\end{tabular}
\label{tab:para}
\end{center}
\vspace{-1.5em}
\end{table}

\begin{table*}[!t]
\renewcommand{\arraystretch}{1.1}
\setlength{\tabcolsep}{8pt}
\begin{center}
\caption{Overall performance of RCKT variants compared with six different baselines. The best result for each metric is in bold. The second-best result is in italics. The best result among the baselines is underlined.}
\begin{tabular}{c|cc|cccccccc}
\hline
\hline
\multirow{2}{*}{Model} & \multicolumn{2}{c|}{Attribute} & \multicolumn{2}{c}{ASSIST09}      & \multicolumn{2}{c}{ASSIST12}      & \multicolumn{2}{c}{Slepemapy}     & \multicolumn{2}{c}{Eedi} \\ \cline{2-11} 
                       & DLKT    & Interpretable    & AUC             & ACC             & AUC             & ACC             & AUC             & ACC             & AUC         & ACC         \\ \hline
DKT                    & $\checkmark$           &                  & 0.7706          & 0.7263          & 0.7287          & 0.7345          & 0.7813          & 0.7988          & 0.7391      & 0.7014      \\
SAKT                   & $\checkmark$           & $\circ^{\mathrm{a}}$                & 0.7674          & 0.7248          & 0.7283          & 0.7344          & 0.7850          & 0.8012          & 0.7417      & 0.7030      \\
AKT                    & $\checkmark$           & $\circ$                & 0.7837          & 0.7343          & \underline{0.7718}          & 0.7536          & 0.7866          & 0.8019          & 0.7828      & 0.7281      \\
DIMKT                  & $\checkmark$           &                  & \underline{0.7854}          & \underline{0.7387}          & 0.7709          & \underline{0.7541}          & \underline{\textit{0.7888}}              &      \underline{0.8021}           &       \underline{0.7835}      &     \underline{0.7285}        \\ \hline
IKT                    &             & $\checkmark$                & 0.7774          & 0.7261          & 0.7624          & 0.7452          &       0.6664          &        0.7846         &      0.7680       &      0.7192       \\
QIKT                   & $\checkmark$           & $\checkmark$                & 0.7815          & 0.7324          &      0.7623           &       0.7462          &        0.7832         &   0.8003              &      0.7803       &    0.7260         \\ \hline
RCKT-DKT              & $\checkmark$           & $\checkmark$                & \textit{0.7929} & \textit{0.7439} & \textit{0.7746} & 0.7545          & 0.7879 & 0.8036          &      \textit{0.7857}       &      \textit{0.7303}       \\
RCKT-SAKT             & $\checkmark$           & $\checkmark$                & 0.7899          & 0.7425          & 0.7728          & \textit{0.7559} & 0.7844          & \textit{0.8041} &       0.7807      &      0.7285       \\
RCKT-AKT              & $\checkmark$           & $\checkmark$                & \textbf{0.7947}*$^\mathrm{b}$ & \textbf{0.7449}* & \textbf{0.7782}* & \textbf{0.7576}* & \textbf{0.7955}* & \textbf{0.8047}* &      \textbf{0.7868}*       &    \textbf{0.7311}*         \\ \hline
improv.                &             &                  & +1.19\%         & +0.83\%         & +0.89\%         & +0.46\%         & +1.13\%         & +0.35\%         &     +0.42\%        &       +0.38\%      \\ \hline\hline
\multicolumn{11}{l}{}\\
\multicolumn{11}{l}{$^{\mathrm{a}}$ $\circ$ means the model leverages the attention mechanism, which is ante-hoc explainable but still lacks interpretability as we discussed before.}\\
\multicolumn{11}{l}{$^{\mathrm{b}}$ * indicates statistical significance over the best baseline, measured by T-test with p-value $\leq$ 0.01.}\\

\end{tabular}
\label{tab:res}
\end{center}
\vspace{-1.5em}
\end{table*}
\section{Experiment}

We design comprehensive experiments to address the following research questions: 
\begin{itemize}[leftmargin=3.2em]
\item[\textbf{Q1:}] Does RCKT successfully address the performance sacrifice issue while improving interpretability?
\item[\textbf{Q2:}] What are the roles and impacts of different components of RCKT in the overall performance and interpretability?
\item[\textbf{Q3:}] How interpretable are the explanations generated by RCKT from the perspective of response influences?
\end{itemize}

In addition, we also conduct experiments on hyper-parameter analysis, knowledge proficiency visualization and response influence approximation analysis to fully demonstrate the effectiveness of RCKT.

\subsection{Experimental Setup}\label{subsec:experimental-setup}
\subsubsection{Datasets} We validate the superiority of RCKT on four widely-used public KT datasets.

\begin{itemize}[leftmargin=*]
\item \textbf{ASSIST09}\cite{b_assist09}\footnote{\url{https://sites.google.com/site/ASSISTmentsdata/home/assistment-2009-2010-data}}. This dataset is collected from ASSISTments, an online tutoring system for mathematics, during the period of 2009 to 2010. We use the \textit{combined} version of this dataset.
\item \textbf{ASSIST12}\cite{b_assist09}\footnote{\url{https://sites.google.com/site/assistmentsdata/home/2012-13-school-data-with-affect}}.
This dataset is also from ASSISTments but during the period of 2012 to 2013.
\item \textbf{Slepemapy}\cite{b_slepemapy}\footnote{\url{https://www.fi.muni.cz/adaptivelearning/data/slepemapy}}.
This dataset is gathered from an online adaptive system called slepemapy.cz, which offers an adaptive practice of geography facts.
We use \textit{place\_asked} as the knowledge concept, and the question identifier is the \textit{type} of questions for each place.
\item \textbf{Eedi}\cite{b_eedi}\footnote{\url{https://eedi.com/projects/neurips-education-challenge}}. This dataset is collected from Eedi, a mathematics homework and teaching platform. It is part of the NeurIPS 2020 education challenge, involving different tasks. We use the \textit{train\_task\_1\_2.csv} file from this challenge as our dataset. Additionally, we use the leaf nodes of the provided math concept tree as the related knowledge concepts for each question.
\end{itemize}

For each dataset, we split every student's response sequence into subsequences of 50 responses each. Subsequences with fewer than 5 responses are removed, and those with fewer than 50 responses are padded with zeros.
The details of the preprocessed datasets are presented in Tab.~\ref{tab:stat}.

\subsubsection{Evaluation}
We adopt a five-fold cross validation to evaluate the performance of the model. In each fold, 10\% of the sequences are used as the validation set for parameter tuning.
As the KT tasks involve binary classification, we utilize the area under the curve (AUC) and accuracy (ACC) as the evaluation metrics.
Additionally, we implement the early stopping strategy during training. This strategy stops the training process when the validation performance does not improve for 10 consecutive epochs.

\subsubsection{Baselines}
We compare RCKT with six baselines, consisting of both DLKT and interpretable methods:

\begin{itemize}[leftmargin=*]
\item \textbf{DKT}~\cite{b_dkt}: A pioneer DLKT method that first applies recurrent neural networks (RNN) for KT.
It outperforms traditional machine learning KT methods.

\item \textbf{SAKT}~\cite{b_sakt}: A DLKT method that first uses the transformer structure in KT. It calculates correlations between target questions and historical responses.

\item \textbf{AKT}~\cite{b_akt}: Another transformer-based DLKT method leveraging the monotonic attention mechanism and the Rasch embedding, which shows strong performance in KT.

\item \textbf{DIMKT}~\cite{b_dimkt}: A state-of-the-art RNN-based DLKT method that fully exploits the question difficulty in KT.

\item \textbf{IKT}~\cite{b_ikt}: An interpretable machine learning KT method employing the tree-augmented naive Bayes classifier (TAN) to model the student response process from three aspects, skill mastery, ability profile, and problem difficulty.

\item \textbf{QIKT}~\cite{b_qikt}: An ante-hoc interpretable DLKT method that employs IRT in the prediction layer from a question-centric level.
\end{itemize}

\subsubsection{Knowledge Encoders}
We extend different knowledge encoders in three DLKT methods bi-directionally to validate the adaptability of RCKT.
\begin{itemize}[leftmargin=*]
\item \textbf{RCKT-DKT}: We use the employed LSTM in DKT and make it bi-directional (BiLSTM).
\item \textbf{RCKT-SAKT}: We employ the transformer knowledge encoder in SAKT but take responses as the query instead of target questions.
\item \textbf{RCKT-AKT}: We apply AKT's monotonic attention in the transformer knowledge encoder as our backbone.
This monotonic attention can also be made bi-directional due to the duality of distance.
\end{itemize}

\subsubsection{Experimental Details}
The experiments are conducted on a Linux server with GTX 2080Ti GPUs. We use the Adam optimizer~\cite{b_adam} to tune all the methods, including RCKT, to their best performance. The learning rate is selected from the set \{5e-3, 2e-3, 1e-3, 5e-4, 2e-4, 1e-4\}, and the dimension size is fixed at 128 for fairness.
All the model-specific hyper-parameters of the baselines are strictly set according to their original papers to ensure a fair comparison.
For RCKT, we list the hyper-parameters in Tab.~\ref{tab:para}.
The values in the braces respectively denote the learning rate, loss balancer $\lambda$, $l_2$ normalization value, dropout ratio, and the number of layers.

\subsection{Overall Performance (Q1)}
Tab.~\ref{tab:res} presents the overall performance of RCKT variants and six other baselines. RCKT-AKT achieves the best performance, outperforming the best baselines by 0.35\% to 1.19\%. 
RCKT with other knowledge encoders also performs competitively.
They obtain seven second places out of eight metrics.
Among the baselines, the conventional DLKT methods, AKT, and DIMKT perform the best due to their powerful neural networks.
The interpretable methods, however, show inferior performance due to the performance sacrifice in achieving interpretability.
Notably, the machine learning method, IKT, demonstrates comparable results on three datasets, but its performance declines on Slepemapy.
This could be attributed to the unbalanced correctness distribution as shown in Table~\ref{tab:stat}.
These results commendably answer the research question that RCKT could also bring performance improvements along with interpretability enhancement.

\begin{table*}[!t]
\renewcommand{\arraystretch}{1.1}
\setlength{\tabcolsep}{4pt}
\begin{center}
\vspace{-.5em}
\caption{Experimental results of the ablation study conducted with the best two encoders DKT and AKT.}
\label{tab:ablation}
\begin{tabular}{l|cccc|cccc|cccc|cccc}
\hline\hline
Dataset & \multicolumn{4}{c|}{ASSIST09}                      & \multicolumn{4}{c|}{ASSIST12}                      & \multicolumn{4}{c|}{Slepemapy}                     & \multicolumn{4}{c}{Eedi}                          \\ \hline
Encoder & \multicolumn{2}{c}{DKT} & \multicolumn{2}{c|}{AKT} & \multicolumn{2}{c}{DKT} & \multicolumn{2}{c|}{AKT} & \multicolumn{2}{c}{DKT} & \multicolumn{2}{c|}{AKT} & \multicolumn{2}{c}{DKT} & \multicolumn{2}{c}{AKT} \\ \hline
Metric  & AUC        & ACC        & AUC         & ACC        & AUC        & ACC        & AUC         & ACC        & AUC        & ACC        & AUC         & ACC        & AUC        & ACC        & AUC        & ACC        \\ \hline
\textbf{RCKT}    & \textbf{0.7929}     & \textbf{0.7439}     & \textbf{0.7947}      & \textbf{0.7449}     & \textbf{0.7746}     & \textbf{0.7545}     & \textbf{0.7782}      & \textbf{0.7576}     & \textbf{0.7879}      & \textbf{0.8036}     & \textbf{0.7955}      & \textbf{0.8047}     & \textbf{0.7857}     & \textbf{0.7303}     & \textbf{0.7868}     & \textbf{0.7311}     \\
-joint  &     0.7894       &      0.7410      &       0.7909      &     0.7413       &    0.7723        &       0.7539     &     0.7756        &     0.7554       &      0.7857      &       0.8014     &      0.7928       &      0.8031      &       0.7823     &      0.7287      &       0.7834     &    0.7292        \\
-mono   &    0.7812        &      0.7311      &     0.7850        &       0.7359     &     0.7691       &       0.7503     &       0.7703      &       0.7522     &      0.7829      &      0.7981      &      0.7901       &      0.7813      &       0.7790     &     0.7259       &      0.7801      &    0.7275        \\
 -con   &      0.7901    &    0.7421        &      0.7918       &      0.7415      &      0.7731      &     0.7540       &       0.7752      &       0.7558             &     0.7853       &      0.8016       &     0.7930       &      0.8033      &      0.7835      &      0.7291      &    0.7841   &   0.7301  \\ 
\hline\hline
\end{tabular}
\end{center} 
\vspace{-1em}
\end{table*}

\subsection{Ablation Study (Q2)}\label{subsec:ablation}

The ablation study aims to evaluate the impact of each component in RCKT by removing specific techniques and comparing the results with the full model. Three components are removed:
\begin{itemize}[leftmargin=*]
\item \textbf{-joint}: The adaptive response probability generator is not jointly trained with the response influence-based counterfactual optimization, which means $\lambda$ is set to 0.
\item \textbf{-mono}: The reliable response retention based on the monotonicity assumption is not performed. Every counterfactual sequence only has an unmasked intervened response.
\item \textbf{-con}: Remove the constraint of response influence to be positive.
\end{itemize}
The ablation study is conducted with DKT and AKT as the encoders, which are the two best-performing ones.
The results in Tab.~\ref{tab:ablation} show that the performance decreases with different degrees of decline when any of the components is removed.
RCKT-mono experiences the most significant deterioration, indicating that the monotonicity assumption is crucial for RCKT's effectiveness.
RCKT-joint also shows a decline, confirming the importance of joint learning with the adaptive response probability generator.
Besides, the degeneration of RCKT-con proves the effect of the constraint on response influences.

\begin{figure}[!t]
\centerline{\includegraphics[width=0.9\linewidth]{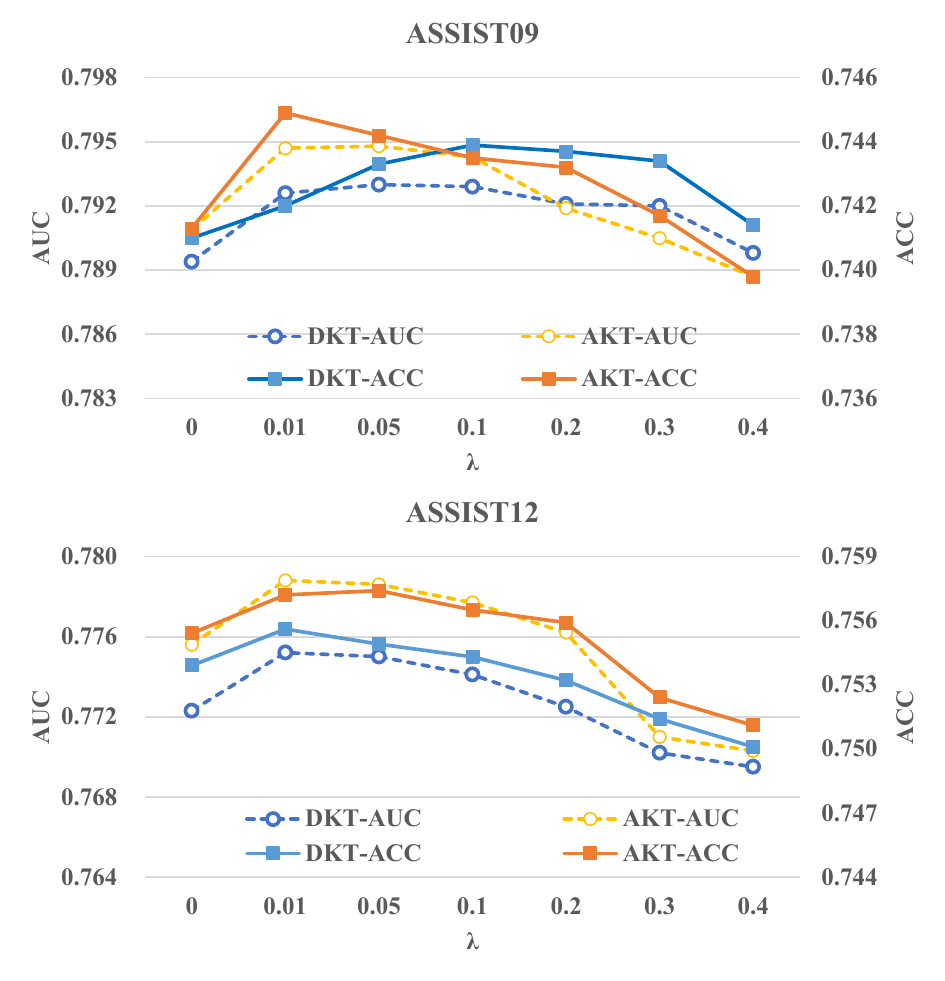}}
\vspace{-1em}
\caption{The performance of RCKT-DKT and RCKT-AKT with different loss balancers on ASSIST09 and ASSIST12.}
\label{fig:lambda}
\vspace{-1em}
\end{figure}

\subsection{Effect of Loss Balancer $\lambda$}
Fig.~\ref{fig:lambda} illustrates the effect of the loss balancer $\lambda$ on the performance of RCKT-DKT and RCKT-AKT on the two ASSIST datasets. The values of $\lambda$ are varied in the range \{0, 0.01, 0.05, 0.1, 0.2, 0.3\}.
The results show that all the metric values reach their peaks in the range of 0.01 to 0.1 for both RCKT-DKT and RCKT-AKT.
This indicates that a proper value of $\lambda$ can help RCKT to be regularized effectively from the joint training, while preventing it from dominating the response influence-based counterfactual optimization.

\begin{figure*}[!t]
\centerline{\includegraphics[width=0.9\linewidth]{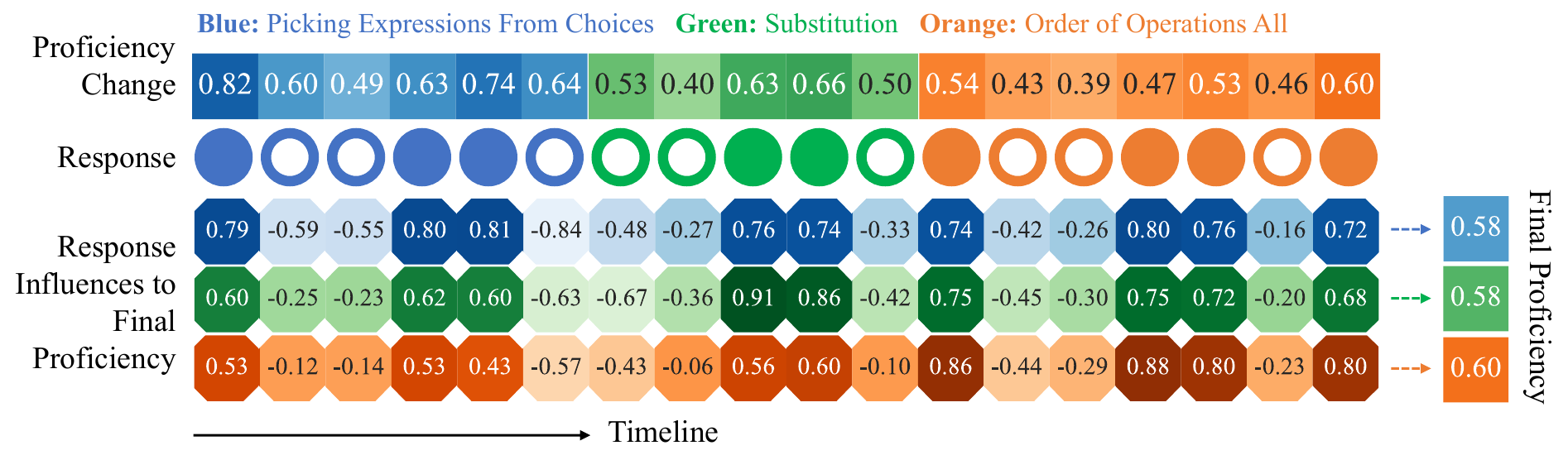}}
\vspace{-1em}
\caption{Interpretable knowledge proficiency tracking of an ASSIST12 student by RCKT.
Each color corresponds to a different knowledge concept as depicted on the top.
The circles represent the student’s responses, with solid ones for correct responses and hollow ones for incorrect responses. The colored octagons on the bottom represent the response influences on capturing the corresponding concepts. 
To make the comparison more obvious, we negate the influences of incorrect responses.
The colored squares on the top indicate the dynamic proficiency of the corresponding concepts after responding to each question, whose values are scaled into $(0,1)$.}
\label{fig:tracing}
\vspace{-.5em}
\end{figure*}

\subsection{Interpretable Knowledge Proficiency Visualization (Q3)}\label{subsec:ikpv}

RCKT also models response influences of capturing knowledge concepts.
A common proficiency tracing approach used in previous works~\cite{b_ekt, b_mrtkt} is to replace the input question ID embeddings with zeros when giving target questions.
We instead use the average ID embeddings of questions related to the traced knowledge concept $k$.
This modifies Eq.~\ref{eq:question_emb} as
\begin{equation}\label{eq:knowledge state tracing}
\textbf{e}=\frac{1}{|\Qset_k|}\sum_{q_i\in\Qset_k}\textbf{q}_i+\textbf{k},
\end{equation}
where $\Qset_k\subset\Qset$ denotes the set of questions related to the concept $k$.
This average manner could make the input subspace more compatible with the question ID embeddings than using zero inputs directly.
In this way, RCKT can track dynamic knowledge proficiency and provide interpretability from response influences.

Quantifying the interpretability of methods based on counterfactual reasoning has two primary approaches: (i) Utilizing data annotations that provide explainable clues~\cite{b_scout}. This approach is unfeasible in current KT scenarios due to the lack of such annotation for explanations. (ii) Removing or solely retaining the generated explainable parts of the input and measuring the output change~\cite{b_cer}. This method is also not applicable because even a minor modification in a student's learning history can lead to differences in their underlying knowledge mastery, affecting other responses, as mentioned earlier (i.e., the first challenge we highlighted). These factors contribute to the challenge of quantifying the interpretability of RCKT. Therefore, we follow the approach of previous KT works~\cite{b_tc_mirt, b_qikt, b_akt}, which conduct case studies to provide an intuitive understanding of interpretability.
Fig.~\ref{fig:tracing} illustrates a case of one student of ASSIST12 answering questions. As shown, the student’s knowledge proficiency increases after he/she has answered a question correctly and decreases otherwise. This is consistent with other KT methods, but for RCKT, these proficiency values are inferred by the response influences. We present the three groups of response influences on mastering different concepts after the student answers the 18 questions at the bottom. Among them, we observe that answering a question related to a concept has a larger (absolute) influence on the proficiency of that concept, e.g., the first six influences for the blue concept. Moreover, the more recent responses have larger influences than the earlier ones, which implies the forgetting behavior. The variation of values across these concepts also suggests the underlying relation between them, i.e., these three concepts all belong to arithmetic. The above observation validates the interpretability of RCKT, which makes the inference process credible.

\begin{figure}[!t]
\centerline{\includegraphics[width=0.96\linewidth]{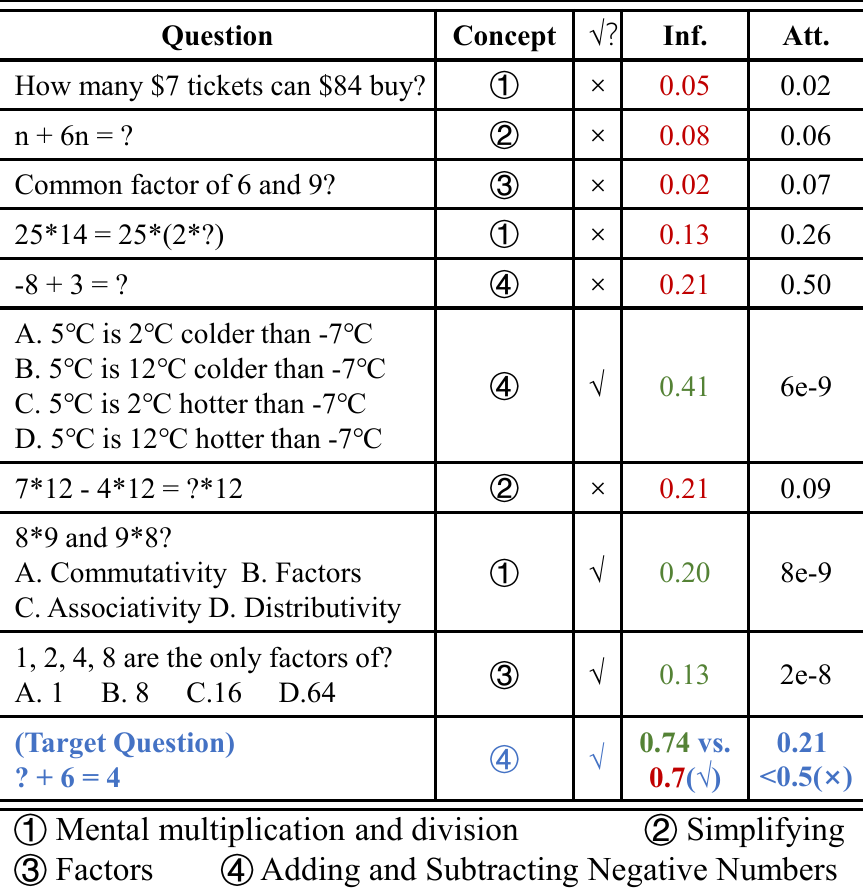}}
\vspace{-.5em}
\caption{A case of predicting one Eedi student performing on a new question given his/her nine historical responses. The green/red values in Inf. denotes the correct/incorrect response influences derived by RCKT-AKT. Att. denotes the average attention values (from multiple heads) generated from SAKT+. The last blue row of these two columns represents their prediction based on the response influences and the attention-fused probability score, respectively.}
\label{fig:interpret}
\vspace{-1em}
\end{figure}

\begin{table}[t]
\begin{center}
\caption{Effeciency comparison of RCKT before and after applying response influence approximation. The average inference time is calculated across all students in the test set.}
\vspace{-1.5em}

\begin{tabular}{l|cc|cc}
\hline\hline
ASSIST09 & \multicolumn{2}{c|}{Before} & \multicolumn{2}{c}{After} \\ \hline
Model             & RCKT-DKT     & RCKT-AKT     & RCKT-DKT    & RCKT-AKT    \\ \hline
AUC             & 0.7896       & 0.7913       & 0.7929      & 0.7947      \\
ACC             & 0.7427       & 0.7434       & 0.7439      & 0.7449      \\
Time/ms           & 214.61       & 305.70       & 10.63       & 14.31       \\ \hline\hline
\end{tabular}
\label{tab:approx}
\end{center}
\vspace{-1.5em}
\end{table}

\subsection{Case of Response Influences (Q3)}
To further demonstrate the interpretability of the response influences in RCKT, we take one Eedi student with his/her 9 responses and a target question as an example.
We use a smaller data file \textit{train\_task\_3\_4.csv} from Eedi that has question texts to help us understand the inference process better.
Fig.~\ref{fig:interpret} shows the example, in which
we compare RCKT-AKT with SAKT+ which is an improved version of SAKT adding question ID embeddings.
Although the student has more incorrect responses than correct ones (6 vs. 3), RCKT still recognizes the influence brought by the correct response to the $6^{th}$ question.
This question shares the same concept as the target question but is more complex. 
This makes the correct response more important and influences the final prediction, which aligns with the ground truth. 
SAKT+, however, pays more attention to incorrect responses and extremely less attention to correct ones. This leads to a wrong prediction, and the inference process is not explicit.
Thus obtaining explanations from these attention values is also non-trivial.

\subsection{Response Influence Approximation Analysis\label{subsec:riaa}}
In this section, we carry out a comparative experiment to demonstrate the effectiveness of RCKT before and after applying the response influence approximation. Due to the large time consumption, we choose the smallest dataset, ASSIST09. 
Similarly, we select the two best encoders, DKT and AKT. As depicted in Table~\ref{tab:approx}, the performance of RCKT exhibits a slight enhancement following the integration of the response influence approximation.
This improvement could be attributed to the utilization of bi-directional encoders. Simultaneously, the inference speed increases by approximately 20 times, validating our earlier theoretical analysis.
\section{Conclusion}
This paper fills the gap in leveraging response influences to improve the interpretability of DLKT models.
The response influences indicate how historical responses affect students answering new questions.
Based on this, we propose RCKT, a framework that uses counterfactual reasoning to quantify the response influences to trace students' knowledge and predict their performance, which is ante-hoc explainable in KT. Comprehensive experiments show that RCKT outperforms most existing KT methods and offers sufficient interpretability.

\end{document}